\pdfoutput=1
\RequirePackage{fix-cm}
\documentclass[english]{article}

\usepackage{amssymb}
\usepackage{amstext}
\usepackage{array}
\usepackage{booktabs}
\usepackage{color}
\usepackage{float}
\usepackage[T1]{fontenc}
\usepackage{fullpage}
\usepackage{graphicx}
\usepackage[latin9]{inputenc}
\usepackage{multicol}
\usepackage[small,raggedright]{titlesec}
\usepackage{units}
\usepackage{url}

\makeatletter

\providecommand{\tabularnewline}{\\}

\floatstyle{ruled}
\newfloat{algorithm}{tbp}{loa}
\floatname{algorithm}{Procedure}
\usepackage[noend]{algorithmic}
\newcommand{\forbody}[1]{ #1 \ENDFOR}
\newcommand{\ifbody}[1]{ #1  \ENDIF}
\newcommand{\whilebody}[1]{ #1  \ENDWHILE}

\PassOptionsToPackage{caption=false}{subfig}
\usepackage{subfig}

\titleformat{\section}
{  \titlerule
   \vspace{.8ex}%
   \large\bfseries\itshape
}  {\thesection.}{.5em}{}

\newenvironment{figurehere}
  {\def\@captype{figure}}
  {}

\makeatother

\usepackage{babel}

\begin{document}

\title{Speculative Parallel Evaluation Of Classification Trees On GPGPU
Compute Engines}

\author{Jason Spencer \thanks{School of Computing and Digital Media, 
   DePaul University, Chicago, IL, USA \newline
   \indent \quad \textbf{email:} jspenc14@cdm.depaul.edu}
}

\date{\today}

\maketitle

\begin{abstract}
We examine the problem of optimizing classification tree evaluation
for on-line and real-time applications by using GPUs. Looking at trees
with continuous attributes often used in image segmentation, we first
put the existing algorithms for serial and data-parallel evaluation
on solid footings. We then introduce a speculative parallel algorithm
designed for single instruction, multiple data (SIMD) architectures
commonly found in GPUs. A theoretical analysis shows how the run times
of data and speculative decompositions compare assuming independent
processors. To compare the algorithms in the SIMD environment, we
implement both on a CUDA 2.0 architecture machine and compare timings
to a serial CPU implementation. Various optimizations and their effects
are discussed, and results are given for all algorithms. Our specific
tests show a speculative algorithm improves run time by 25\% compared
to a data decomposition.

\textbf{keywords:} Classification Trees, Decision Tree Evaluation,
Parallel Algorithms, GPU Computing, Speculative Decomposition,
Optimization, Image Segmentation.
\end{abstract}

\begin{multicols}{2}

\section{\label{sec:Intro}Introduction}

Classification trees are used to solve problems in areas as diverse
as target marketing, fraud detection, pattern recognition, computer
vision, and medical diagnosis. In many applications, classification
trees are carefully designed once but then applied to many data sets
to provide automated classifications. This approach is used to create
validated classifiers for tissue classification in mammography \cite{Oliver05}
and intravascular ultrasound \cite{Nair02} diagnostic procedures.
While training the classifier is done offline, tree evaluation of
each patient's data in these applications is an on-line algorithm
where a user waits for a classification to be performed on many, many
samples. Time spent waiting for this evaluation consumes valuable
procedure room equipment and personnel. Performance requirements only
increase when single images are replaced by moving video for computer
vision applications, as in robotic navigation \cite{Baumstarck}.
In this environment, many classifications are needed in real-time
to compute and affect a timely response. Thus the need for high-performance
on-line evaluation of classification trees ranges from beneficial
to absolutely necessary. 

The assignment of a class to a given sample from a dataset requires
that the sample be evaluated at each decision point along its path
from the root of the tree to its eventual terminal leaf. While it
may seem that each decision must be made in series for that sample,
we note that each sample's classification is independent of all other
samples. This allows us to decompose the problem of classifying all
samples in a dataset into the independent problems of classifying
each sample, which can be done in parallel. This decomposition according
to sample data (a data decomposition approach) makes a growing number
of parallel computing architectures available to speedup tree evaluation. 

There is a good deal of literature on parallelization of training
algorithms used to create classification trees \cite{Ben-Haim2010,Jin2002,Scalparc,Sliq,Sprint,Zaki98}
in a traditional parallel processing setting. Research on the tree
evaluation problem, however, seems to focus on Graphics Processing
Units (GPUs) as the implementation platform. GPUs are typically designed
specifically for data parallel applications. As inexpensive, commodity
hardware found on every standard PC, GPUs match the cost, size, and
power requirements of the on-line tree evaluation problem setting
more closely than traditional super computers. Such application of
graphics hardware to generic problems has become known as General
Purpose GPU (GPGPU) computing.

An early expedition into GPGPU techniques for machine learning can
be found in \cite{GpuMl}, but application to tree evaluation was
first proposed by Sharp in \cite{Sharp08}. His framework stores
the tree as an array of nodes containing the decision criteria of
that node and an index used to locate the next node. Subsequent node
indices are computed without conditional branches to avoid their heavy
performance penalties on most GPUs. The tree definition is passed
to the GPU as a texture map used by a custom pixel shader. The shader
consumes input feature data and combines it with the texture to produce
a final value, the assigned class, for each pixel in parallel. Sharp
extends this to evaluate random forests by concatenating multiple
tree structures in the texture data and iterating over all trees.
Results show a speedup of roughly two orders of magnitude over host-based
algorithms.

In \cite{Baumstarck}, Baumstarck also uses a data parallel approach
on GPUs for a computer vision application, available in \cite{Stair}.
The implementation is done directly on the Compute Unified Device
Architecture (CUDA) platform offered by NVIDIA Corporation \cite{CudaZone}
without using graphics libraries. Though conditionals are used in
the tree traversal, Baumstarck reports a fifty-fold speedup of forest
evaluation.

In this paper, we investigate a speculative approach to tree evaluation
on massively parallel GPU architectures, namely CUDA. Rather than
treating the full evaluation of one sample as the atomic parallel
task, we parallelize the evaluation of each node in the tree for a
single sample then reduce the resulting path through the tree in parallel.
This approach has some performance benefits on architectures where
execution of parallel processors is not independent, as in SIMD machines.
We compare this approach to the data decomposition used in previous
work and to the best-known serial host algorithm, both of which we
restate here so that all approaches are put on a solid footing. In
the specific environment we examine, results for speculative decomposition
show a 25\% performance improvement over data decomposition. We also
see that host memory bandwidth and data distribution is an important
measurement consideration that can dominate the nuances of GPU performance
gains in typical PC systems, and must be accounted for in any statement
of speedup results.

\section{\label{sec:Prelims}Preliminaries}

\subsection{\label{sub:ClassificationTrees}Classification Trees}

In evaluating a classification tree, we are given a set of records,
called the \emph{dataset}, and a full binary decision tree, called
the \emph{classifier}. Each record in the dataset contains several
fields, called \emph{attributes} or \emph{features}. One of the attributes,
the \emph{classifying} attribute, indicates to which \emph{class}
the record belongs and is unknown. In the general case, attributes
can be \emph{continuous}, having (real) numerical values from an ordered
domain, or \emph{categorical}, representing values from an unordered
set. The classifier is a predictive model created through a process
known as \emph{training}. In training, observations on a \emph{training
set} of records, each having a known classifying attribute, are used
to build a tree such that each interior, or \emph{decision}, node
uses a single attribute value test to partition the set of records
recursively until the subset of records at a given node have a uniform
class. Such nodes are encoded in the tree as leaf nodes. The evaluation
of a dataset is complete when the trained classifier is used to determine
to which leaf, and thereby which class, each record belongs.

There are several training algorithms for examining attributes and
generating trees. The particular algorithm used will not concern us
here, so long as the resulting tree has the above properties. We examine
trees where all attributes are continuous, a common occurrence in
image segmentation. While we will look at real-valued attributes (approximated
with floating point numbers), ordered discrete values would behave
very much the same. Categorical attributes, though, would likely require
some modifications to our approach. We will further assume that class
values can be enumerated and put into one-to-one correspondence with
the natural numbers. Evaluation will operate only on numbers, and
any mapping to another representation for class values (e.g. to descriptive
strings or pixel values) will be done outside the evaluation process.

\subsection{\label{sub:CudaGpus}CUDA GPUs}

GPGPU computing has grown in popularity in recent years as a technique
for improving performance for massively parallel applications, especially
where visualization and images are concerned. Initially, generic parallel
computing was achieved on GPUs by cleverly mapping the processing
into the graphics domain using libraries such as OpenGL to perform
primitive tasks. As demand for customized graphics processing grew,
vendors began supporting domain-specific programming languages like
GL Shading Language (GLSL), making the GPU's floating point units
more available. 

In recent years, GPGPU computing frameworks have made great strides
in removing assumptions about the domain and providing a generic capability
to be used in any application needing massive parallelization. Perhaps
the leading such framework, NVIDIA's CUDA architecture, can add tens
or hundreds of GigaFLOPs to a system's capability on a single adapter
card. 

This power can be brought to bear on generic problems with great ease
of use. The programming environments for these devices, whether vendor-specific
or the industry standard OpenCL, can be used with no reference to
the graphics domain. These environments subset the C/C++%
\footnote{NVIDIA represents that CUDA is a extension to ANSI C, but recent versions
also allow for the use of classes.%
} programming language and provide a set of keyword extensions to manage
the generation of both device-specific code and host code from the
same source file set. In this way, code written to run on the GPU,
called a \emph{kernel}, is invoked with something that feels very
akin to a C function call.

\subsubsection{\label{subsub:CudaProgramming}CUDA Programming Model}

The CUDA runtime executes kernels across many \emph{threads}, or individual
streams of instructions (usually for a single atomic parallel task),
and manages the mechanics of scheduling in hardware. Threads are grouped
into \emph{blocks} as 1, 2, or 3 dimensional arrays with each thread
having a unique identifying index in each dimension of the block.
Further, blocks are grouped into a 1 or 2 dimensional \emph{grid},
with each block again having an identifying index in the grid dimensions.
Each kernel invocation is done over a single grid and gives the grid
and block dimensions to use when launched. Threads within a block
are allowed to synchronize and share memory, but no communication
between blocks is allowed. Threads are scheduled and executed in 32-thread
units called a \emph{warp}, with some operations happening on a \emph{half-warp},
or 16 threads. All threads have access to their local memory (registers
and stack), the shared memory of their block, and a global memory
common to the entire device. The host can read from and write data
to global memory but not local or shared memory. The host is required
to copy kernel input and output data to and from device global memory
outside of the kernel execution.

A simple example helps to illustrate a typical kernel invocation.
First, the host CPU copies the input data to the GPU device's global
memory. Since the host and device address spaces are separate, the
CUDA runtime provides the host with APIs to allocate storage in device
space, copy memory between spaces, look up device space symbol addresses,
etc. The host must also allocate device global memory to store the
results of the computation. The host can then invoke the CUDA runtime
to launch the kernel with certain grid and block dimensions. Arguments
such as the input and output buffers in device space are passed in
the invocation. The device allocates execution resources to the kernel
grid and schedules threads to execute in warps. Each thread uses its
block and thread indices to identify its associated portions of input
and output data. It can then do thread-specific memory transfer to
its own stack and registers. Once the input data is locally available,
computation is done and output is stored in device global memory.
When all threads have completed, the host is signaled and is then
free to copy the results from device to host memory and deallocate
buffers.

\subsubsection{\label{subsub:CudaHW}CUDA Hardware Architecture}

While an extensive discussion of CUDA architecture is beyond the scope
of this paper, some of the algorithm designs given herein are driven
by certain qualities which bear discussion. The fundamental execution
units of a CUDA device, called \emph{stream processors} and known
as \emph{cores}, are arranged in $N$-way SIMD groups for some implementation-dependent
$N$ (usually 8, 32, or 48). These groups are combined with super
function units (SFUs), instruction cache/decode logic, a register
file, L1 cache/shared memory, (usually 2) warp schedulers, and a network
interconnect to form a \emph{streaming multiprocessor}, or SM (Figure
\ref{fig:SM-detail}). All threads in a block will be executed on
the same SM, scheduled very efficiently by the hardware warp schedulers.
When a warp is scheduled, all threads in that warp execute the same
instruction, but have their own registers and stack. When some threads
take conditional branches different from other threads, the warp executes
the two paths in series until the paths merge. This is known as a
\emph{divergent path}, and can affect the kernel's performance substantially.

When a warp encounters a long-latency instruction (such as global
memory access), it can be swapped for another warp in a small number
of clocks. There is a limit to this capability, however, and the SM
can only have so many blocks and threads resident at a time. This
concept is known as \emph{occupancy}, and can also affect the kernel's
performance. Low occupancy means an SM has nothing to do during long
latency instructions, so the SM is not fully utilized.

\begin{figurehere}
   \begin{centering}
      \includegraphics[height=12cm]{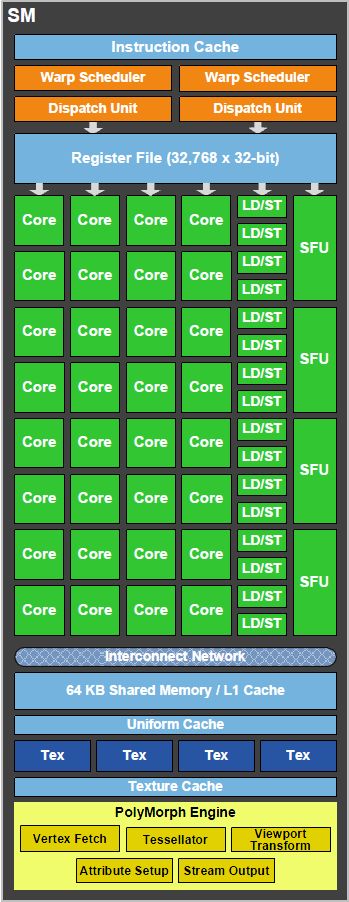}
      \par
   \end{centering}

   \caption{\label{fig:SM-detail}Streaming Multiprocessor detail (NVIDIA Corporation)}
\end{figurehere}

Finally, accessing global memory from an SM is an expensive operation,
typically 100 times the cost of accessing local memory. In some CUDA
implementations, accesses to global memory that meet certain requirements
(such as contiguous access of 32, 64, or 128 bytes made in order by
each core) can be \emph{coalesced} into a single read, improving throughput.
Later versions of CUDA hardware add L1 and even L2 cache, which mitigates
the cost of non-coalesced reads.

See \cite{CudaZone,DevZone,Kirk10,Sanders2010CUDA} for a more complete
and detailed overview of the CUDA architecture.

\section{\label{sec:ClassificationAlgs}Classification Tree Algorithms}

It is natural to imagine an algorithm for evaluating a decision tree
using a binary tree data structure and a depth-first traversal which,
at each node, uses a conditional to evaluate whether the traversal
should follow the left or right child of the node. Conditional statements,
however, present problems for traditional CPUs (in the form of branch
misprediction and pipeline flush) and GPUs (in the form of serialized
divergent paths for SIMD warp execution.) Sharp avoids this problem
in \cite{Sharp08} by developing a branchless tree traversal, which
we will adopt for the base serial evaluation algorithm. A host implementation
of this algorithm, as the best known serial algorithm, will be the
reference by which speedup of parallel algorithms is determined.

\subsection{\label{sub:BranchlessEval}Branchless Tree Evaluation}

The evaluation problem can be stated as follows: given a dataset $\mathcal{D}=\{R:R=(r_{1},\ldots,r_{A}),\: r_{a}\in\mathbb{R}\}$
with $\mid\!\mathcal{D}\!\mid=M$ and a full binary classification
tree $\tau$ with a set of nodes $\mathcal{N}=\{n:n=(a_{n},t_{n},d_{n}^{\, r},d_{n}^{\, l},c_{n})\}$
where:
\begin{itemize}
\item $\mid\!\mathcal{N}\!\mid=N$ is the number of nodes in $\tau$
\item $1\leq a_{n}\leq A$ is the index of attribute $r_{a_{n}}$ in each
record $R$ to be evaluated by node $n$
\item $t_{n}\in\mathbb{R}$ is the threshold for attribute $r_{a_{n}}$
used by node $n$ 
\item $d_{n}^{\, l}\in\{\mathcal{N\,}\bigcup\emptyset\}$ is $n$'s left
descendant and recursively evaluates $R$ when $r_{a_{n}}\leq t_{n}$
\item $d_{n}^{\, r}\in\{\mathcal{N\,}\bigcup\emptyset\}$ is $n$'s right
descendant and recursively evaluates $R$ when $r_{a_{n}}>t_{n}$
\item $c_{n}\in\{\mathcal{C\,}\bigcup\bot:\:\mathcal{C}\subset\mathbb{N}$ is 
the set of possible class values$\}$ is $\bot$ when 
$(d_{n}^{\, r}\neq\emptyset\,\bigwedge\, d_{n}^{\, l}\neq\emptyset)$ or 
some $c\in\mathcal{C}$ when 
$(d_{n}^{\, r}=\emptyset\,\bigwedge\, d_{n}^{\, l}=\emptyset)$ 
\end{itemize}
and having a root node $n_{0}$, assign to each $R\in\mathcal{D}$
a $c_{R}\in\mathcal{C}$ by recursively evaluating $R$ beginning
at $n_{0}$. 

\begin{algorithm}[H]
\begin{algorithmic}
[1]\STATE{$breadthFirstTree=[\,]$}
\STATE{$Q$=queue()}
\STATE{push($Q$, $n_{0}$)}
\STATE{$i$= 0}
\STATE{$childIndex$= 1}

\WHILE{$Q$ not empty}
\whilebody{%
   \STATE{$n$ = pop($Q$)}
   \STATE{$node$.attributeIndex = $a_{n}$}
   \STATE{$node$.threshold = $t_{n}$}
   \STATE{$node$.classVal = $c_{n}$}
   \STATE{$node$.child = $childIndex$}
   \STATE{$breadFirstTree${[}$i${]} = $node$}
   \STATE{$i=i+1$}
   \IF{$d_{n}^{\, l}\neq\emptyset$}
   \ifbody{%
      \STATE{push($Q$, $d_{n}^{\, l}$)}
      \STATE{$childIndex=childIndex+1$}
   }
   \IF{$d_{n}^{\, r}\neq\emptyset$}
   \ifbody{%
      \STATE{push($Q$, $d_{n}^{\, r}$)}
      \STATE{$childIndex=childIndex+1$}
   }
}
\end{algorithmic}

\caption{\label{alg:BreadthFirstEnc}Breadth-first Encoding of Tree}

\end{algorithm}

\begin{algorithm}[H]
\begin{algorithmic}
[1]\STATE{\textbf{Parameter:} $\mathcal{D}$}
\STATE{\textbf{Parameter:} $breadthFirstTree[N]$}
\STATE{\textbf{Output:} $assignedClasses[\mid\!\mathcal{D}\!\mid]$}
\FORALL{$R\in\mathcal{D}$}
\forbody{
   \STATE{$i=0$}
   \WHILE{$breadthFirstTree[i].\text{classVal}=\bot$}
   \whilebody{
      \STATE{$a=breadthFirstTree[i].\text{attributeIndex}$}
      \STATE{$t=breadthFirstTree[i].\text{threshold}$}
      \STATE{$i=breadthFirstTree[i].\text{childIndex}+(r_{a}>t)$}
   }
   \STATE{$c_{R}=breadthFirstTree[i].\text{classVal}$}
   \STATE{$assignedClasses[R]=c_{R}$}
}
\end{algorithmic}

\caption{\label{alg:SerialEval}Serial Tree Evaluation}

\end{algorithm}

To evaluate $\tau$ without branching, we first encode $\mathcal{N}$
in a breadth-first array of nodes. Procedure \ref{alg:BreadthFirstEnc}
shows how each node is assigned an index $i$ in the array $breadthFirstTree$
to create a data structure describing the tree. Note that every right
child has an index that is one more than the neighboring left child.
Each node, then, need only store the index of its left child. To compute
the index of the next node to evaluate, the node compares its attribute
value $r_{a_{n}}$ against its threshold $t_{n}$ using the Boolean
predicate {}``greater-than.'' If the result is false and encoded
as 0, adding the result to the node's child index will yield the index
of its left child, as desired. If the result is true encoded as a
1, adding it to the child index will yield the node's right child's
index. While not strictly branchless due to the while loop, this technique
does avoid any explicit conditional to compute the path to take at
each decision node. The general algorithm is shown in Procedure \ref{alg:SerialEval}.

\subsection{\label{sub:DataDecomposition}Data Decomposition}

Procedure \ref{alg:SerialEval} is parallelized by data decomposition
almost trivially, since each record is independent of the others.
We can simply assign $m$ records to $p$ processors and have each
loop only over $m$. The only additional work is to map the $m$ records
to the global dataset for the purposes of indexing into the input
and output arrays. Procedure \ref{alg:DataEval} shows the algorithm
for each processor with indexing details for parameters $\mathcal{D}$
and $assignedClasses$. We use $\mathcal{D}[s..t)$ to mean the subset
of elements of $\mathcal{D}$ beginning at element $s$ up to but
not including element $t$. Here, we assume a shared memory architecture
so that all processors have equal access to the parameter and output
buffers. Knowing the index to a record $R$ in $\mathcal{D}$ also
gives the index to the corresponding $assignedClasses$ value. The
steps of making $\mathcal{D},\; breadthFirstTree,\text{ and }assignedClasses$
available to each processor are omitted.

\cite{Sharp08} uses a data parallel approach similar to this, as
does \cite{Baumstarck} when evaluating boosted decision trees using
CUDA, though the later uses conditional instructions to traverse the
tree.

\begin{algorithm}[H]
\begin{algorithmic}
[1]\STATE{\textbf{Parameter:} $\mathcal{D}$}
\STATE{\textbf{Parameter:} $breadthFirstTree[N]$}
\STATE{\textbf{Parameter:} $m\in\mathbb{N}$, the number of records for this
processor to process}
\STATE{\textbf{Parameter:} $p\in\mathbb{N}$, this processor's rank}
\STATE{\textbf{Output:} $assignedClasses[\mid\!\mathcal{D}\!\mid]$}
\FORALL{$R\in\mathcal{D}[m\cdot p\,..\, m(p+1))$}
\forbody{
   \STATE{$i=0$}
   \WHILE{$breadthFirstTree[i].\text{classVal}=\bot$}
   \whilebody{
      \STATE{$a=breadthFirstTree[i].\text{attributeIndex}$}
      \STATE{$t=breadthFirstTree[i].\text{threshold}$}
      \STATE{$i=breadthFirstTree[i].\text{childIndex}+(r_{a}>t)$}
   }
   \STATE{$c_{R}=breadthFirstTree[i].\text{classVal}$}
   \STATE{$assignedClasses[R]=c_{R}$}
}
\end{algorithmic}

\caption{\label{alg:DataEval}Data-Parallel Tree Evaluation}

\end{algorithm}

\subsection{\label{sub:SpeculativeDecomposition}Speculative Decomposition}

While a data decomposition applies multiple processors to the evaluation
problem very efficiently, the task of evaluating a single tree is
still done serially. Once $m$ is reduced to 1, no further processors
can be applied to the problem usefully. Also, very deep and unbalanced
trees may lead to asymmetries in the runtime between processors. In
image segmentation, for instance, neighboring samples are expected
to take similar paths through the tree and have almost uniform class
values. By luck of the draw, some processor may be assigned $m$ records
that happen to be classified by the deepest node in the tree while
others have records classified at the top of the tree. This leads
to idle time in the {}``lucky'' processors, and thereby, practical
inefficiency. Further, adjacent records taking different paths leads
to similar inefficiencies in SIMD architectures like CUDA SMs or Intel's
SSE instruction set.

We propose a speculative decomposition of the problem to avoid the
issues of divergent paths, irregular memory access patterns, idle
time due to asymmetrical processing times, and to provide more uniform
evaluation times needed in deterministic, real-time applications.
We assign to each record a group of $p$ processors, called a \emph{record
group}, such that $p=N$. If there are $G$ such groups, the total
number of processors becomes $P=Gp$. Within the group, each node
$n$ of the tree is assigned to processor $p_{n}$. The first step
of the algorithm is to evaluate all nodes in parallel. Each processor
stores the child node index $i$ determined by the node evaluation
into a shared memory array, $path$, having one element for each processor.
The second step is to reduce the path through the tree to the selected
leaf node. This is done by having each processor copy the $path$
value of its child node into its own element of $path$. That is,
each node finds its successor's successor and adopts that as its own
successor. We can then think of the $path$ array as storing the eventual
successor for each node, with the eventual successor of the root node
being the terminal node for the record. This step must be done synchronously
across all processors in the record group. Leaf nodes are specifically
designed to always evaluate to themselves by setting their threshold
to $-\infty$ and their child index to be their own index.

\begin{figure*}
  \centering
   \begin{minipage}[c][1\totalheight][t]{0.60\columnwidth}%
      \centering
         \subfloat[Example Tree]{%
            \includegraphics[clip,scale=0.75]{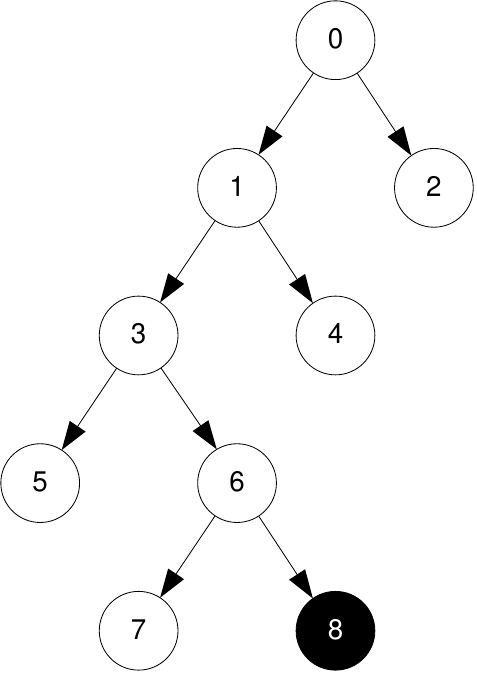}
         }%
         \par
   \end{minipage}
   \begin{minipage}[c][1\totalheight][t]{0.70\columnwidth}%
      \centering
         \subfloat[\label{fig:path-after-Node}$path$ after Node Evaluation]{%
            \includegraphics[clip,scale=0.75]{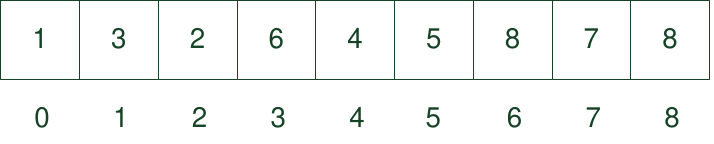}
         }%
         \par
         \subfloat[\label{fig:path-after-One}$path$ after One Reduction Iteration]{%
            \includegraphics[clip,scale=0.75]{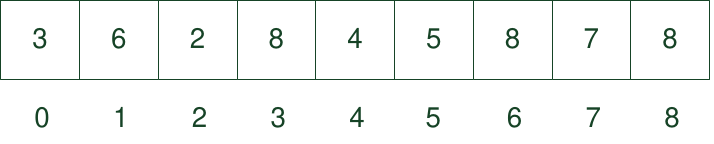}
         }%
         \par
         \subfloat[\label{fig:path-after-Two}$path$ after Two Reduction Iterations]{%
            \includegraphics[clip,scale=0.75]{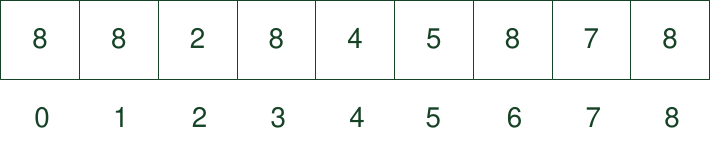}
         }%
         \par
   \end{minipage}
   \caption{\label{fig:Parallel-Tree-Reduce}Parallel Tree Path Reduction}
\end{figure*}

Figure \ref{fig:Parallel-Tree-Reduce} shows an example tree and the
$path$ array after the initial node evaluation (\ref{fig:path-after-Node}),
then after one (\ref{fig:path-after-One}) and two (\ref{fig:path-after-Two})
steps of the parallel reduction phase. Note that for a tree of depth
$d$, only $\Theta(\log_{2}d)$ reduction steps are necessary for
the root node to arrive at the terminal leaf's index. When this occurs,
the reduction terminates. 

Procedure \ref{alg:SpecEval0} gives the parallel algorithm, which
handles indexing the dataset as before but now accounts for the specific
record group $g$ in the calculation as well as determining which
node of the tree each processor is assigned to and setting up the
shared variable $path$. To compute the dataset indices, we can follow
the form of Procedure \ref{alg:DataEval} but substitute $g$ for
$p$. Again, we assume a shared arrangement for the input $dataset$
and the output $assignedClasses$ where the indices in each array
correspond naturally. We use the primitive \textbf{barrier}() to provide
synchronization on updates to $path$ from within record group $g$.

\begin{algorithm}[H]
\begin{algorithmic}
[1]\STATE{\textbf{Parameter:} $\mathcal{D}$}
\STATE{\textbf{Parameter:} $breadthFirstTree[N]$}
\STATE{\textbf{Parameter:} $m\in\mathbb{N}$, the number of records for this
record group to process}
\STATE{\textbf{Parameter:} $g\in\mathbb{N}$, the record group this processor
belongs to }
\STATE{\textbf{Parameter:} $p_{n}\in\mathbb{N}$, this processor's rank in
the record group}
\STATE{\textbf{Output:} $assignedClasses[\mid\!\mathcal{D}\!\mid]$}
\STATE{\textbf{Shared Variable:} $path[N]$}
\FORALL{$R\in\mathcal{D}[m\cdot g\,..\, m(g+1))$}
\forbody{
   \STATE{$a=breadthFirstTree[p_{n}].\text{attributeIndex}$}
   \STATE{$t=breadthFirstTree[p_{n}].\text{threshold}$}
   \STATE{$path[p_{n}]=breadthFirstTree[p_{n}].\text{childIndex}+(r_{a}>t)$}
   \STATE{\textbf{barrier}($g$)}
   \STATE{$rootClass=$\\ \quad $breadthFirstTree[path[0]].\text{classVal}$}
   \WHILE{$rootClass=\bot$}\label{Flo:WhileClassSpecEval0}
   \whilebody{
      \STATE{$path[p_{n}]=path[path[p_{n}]]$}
      \STATE{\textbf{barrier}($g$)}
      \STATE{$rootClass=$\\ \quad $breadthFirstTree[path[0]].\text{classVal}$}
   }
   \STATE{$c_{R}=rootClass$}
   \STATE{$assignedClasses[R]=c_{R}$}
}
\end{algorithmic}

\caption{\label{alg:SpecEval0}Speculative Parallel Tree Evaluation}

\end{algorithm}

\subsection{\label{sub:ImprovedSpeculativeDecomposition}Improved Speculative
Decomposition}

A few inefficiencies exist in Procedure \ref{alg:SpecEval0}. First,
processors assigned to leaf nodes will always produce the same, known
output, and so their assigned processors do no productive work. To
avoid this waste, the $path$ array can be initialized with the known,
static results for all leaves. Processors will only be assigned to
decision nodes such that $0\leq p_{n}<(N-1)/2$, the number of internal
nodes in a full binary tree. This means, however, that mapping processors
in a record group to tree nodes is no longer a simple, sequential
operation. A tree-specific look-up table can accommodate this. As
the record group processes, each processor will modify only the element
of $path$ it is assigned to. 

Second, if the tree reduction is viewed probabilistically, we see
that most records will end up at some leaf between levels 1 and $d$
of the tree, averaging to some $d_{\mu}$ for the dataset. Checking
the while condition on line \ref{Flo:WhileClassSpecEval0} of Procedure
\ref{alg:SpecEval0} for all levels $d_{r}<d_{\mu}$ leads to an expected
inefficiency. If $d_{\mu}$ is known or can be determined experimentally
for the tree, reducing $d_{\mu}$ levels in a single while loop pass
can provide an average case performance enhancement by reducing loop
iterations and the number of barrier operations required.

Procedure \ref{alg:SpecEval1} gives the improved parallel algorithm
for speculative decomposition. We add the static paths for the leafs
of the tree on line \ref{Flo:leafPaths}, and use that to initialize
the $path$ array in parallel on line \ref{Flo:initPath}. Each processor
must now initialize two elements of $path$ since there are only processors
for the internal nodes. We also add the processor-node map on line
\ref{Flo:nodeMap}, which records the node index $i$ assigned to
each processor. Line \ref{Flo:pathIndexing} shows the concept of
multiple reductions per loop, though the optimal implementation will
be tree-specific.

\begin{algorithm}[H]
\begin{algorithmic}
[1]\STATE{\textbf{Parameter:} $\mathcal{D}$}
\STATE{\textbf{Parameter:} $breadthFirstTree[N]$}
\STATE{\textbf{Parameter:} $leafPaths[N]$}\label{Flo:leafPaths}
\STATE{\textbf{Parameter:} $processorNodeMap[(N-1)/2]$}\label{Flo:nodeMap}
\STATE{\textbf{Parameter:} $m\in\mathbb{N}$, the number of records for this
record group to process}
\STATE{\textbf{Parameter:} $g\in\mathbb{N}$, the record group this processor
belongs to }
\STATE{\textbf{Parameter:} $p_{n}\in\mathbb{N}$, this processor's rank in
the record group}
\STATE{\textbf{Output:} $assignedClasses[\mid\!\mathcal{D}\!\mid]$}
\STATE{\textbf{Shared Variable:} $path[N]$}
\STATE{$path[2p_{n}]=leafPaths[2p_{n}]$}\label{Flo:initPath}
\STATE{$path[2p_{n}+1]=leafPaths[2p_{n}+1]$}
\STATE{$i=processorNodeMap[p_{n}]$}
\FORALL{$R\in\mathcal{D}[m\cdot g\,..\, m(g+1))$}
\forbody{
   \STATE{$a=breadthFirstTree[i].\text{attributeIndex}$}
   \STATE{$t=breadthFirstTree[i].\text{threshold}$}
   \STATE{$path[i]=breadthFirstTree[i].\text{childIndex}+(r_{a}>t)$}
   \STATE{\textbf{barrier}($g$)}
   \STATE{$rootClass=$\\\quad $breadthFirstTree[path[0]].\text{classVal}$}
   \WHILE{$rootClass=\bot$}
   \whilebody{
      \STATE{$path[i]=path[path[path[i]]]$}\label{Flo:pathIndexing}
      \STATE{\textbf{barrier}($g$)}
      \STATE{$rootClass=$\\\quad $breadthFirstTree[path[0]].\text{classVal}$}
   }
   \STATE{$c_{R}=rootClass$}
   \STATE{$assignedClasses[R]=c_{R}$}
}
\end{algorithmic}

\caption{\label{alg:SpecEval1}Speculative Parallel Tree Evaluation}

\end{algorithm}

\subsection{Management and Tuning of Parallel Algorithms}

Some management work is required for each algorithm in sections 
\ref{sub:DataDecomposition}, \ref{sub:SpeculativeDecomposition}, 
and \ref{sub:ImprovedSpeculativeDecomposition}, but is omitted for 
brevity and to preserve generality. This includes making the buffers 
for $\mathcal{D}$, $assignedClasses$, $breadthFirstTree$, and any of 
the other necessary symbols available to all the parallel processors 
for each algorithm. The mechanism for sharing these buffers depends 
on the programming environment used. Also, selection of optimal values 
for $G$ and $m$ given $P$, $N$, $M$, and the available execution 
hardware architecture is critical but entirely implementation dependent.

\subsection{Analysis of Evaluation Algorithms}

We now analyze the asymptotic behavior of these general algorithms
assuming a traditional parallel processing setting of independent
processors connected via shared memory. We perform an average case
run time analysis by assigning $d_{\mu}$to the be average depth of
the tree traversed by the records in the dataset. This can be determined
if the entire dataset is known \emph{a priori}, or can be statistically
estimated given an significant sample size, such as the training set.
The serial runtime for Procedure \ref{alg:SerialEval} for $M$ records
is given by \[T_{2}=Md_{\mu}(t_{e}+t_{c})\]

\noindent where $t_{e}$ is the time to evaluate a node's attribute against
its threshold and $t_{c}$ is the time to compare the new node's class
value to $\bot.$ We also refer to $t_{n}=t_{e}+t_{c}$ as the time
needed to evaluate a node.

The run time for Procedure \ref{alg:DataEval} is a function of $P$,
the total number of processors applied, and is given by 

\[
T_{3}(P)=\frac{M}{P}d_{\mu}(t_{e}+t_{c})+t_{i}+t_{s}(M)\]

\noindent where each processor classifies $\frac{M}{P}$ records, $t_{i}$ is
the time needed to compute the index in $\mathcal{D}$ assigned to
the each processor, and $t_{s}(M)$ is the time needed to transmit
$M$ records on the shared memory machine for processing. We can then
examine the speedup of Procedure \ref{alg:DataEval} as
\begin{eqnarray*}
S_{3}(P)=\frac{T_{2}}{T_{3}(P)} & = & \frac{Md_{\mu}(t_{e}+t_{c})}{\frac{M}{P}d_{\mu}(t_{e}+t_{c})+t_{i}+t_{s}(M)}\\
 & = & \frac{P}{1+\frac{P\left(t_{i}+t_{s}(M)\right)}{Md_{\mu}(t_{e}+t_{c})}}
\end{eqnarray*}

\noindent If we assume $t_{s}(M)=\sigma M+\gamma$ for some $\sigma,\gamma$
and we ignore $\gamma\text{ and }t_{i}$ as small constants, then
this simplifies asymptotically to \[
S_{3}(P)\approx\frac{P}{1+\frac{P\sigma}{d_{\mu}t_{n}}}\]
 which suggests the speedup will be decided by the relative performance
of the memory copy and the serial node processing time. If they are
very similar, we would not expect much speedup. If memory copies are
very fast compared to node processing, some benefit may be had. Likewise
for the efficiency, given by 

\begin{eqnarray*}
E_{3}(P)=\frac{S_{3}(P)}{P} & \approx & \frac{1}{1+\frac{P\sigma}{d_{\mu}t_{n}}}
\end{eqnarray*}

\noindent we expect good results only when copy time is much less than processing
time.

For Procedure \ref{alg:SpecEval1}, the analysis is a bit more involved.
If each group of processors is assigned $m=\frac{M}{G}$ records for
$G$ groups of $p$ processors such that $P=Gp$, the parallel runtime
is given by \[
T_{5}(P)=\frac{Mp}{P}\left(t_{e}+\left(\log_{2}d_{\mu}\right)t_{c}\right)+t_{i}+t_{s}(M)\]

\noindent and the speedup is 
\begin{eqnarray*}
S_{5}(P) & = & \frac{T_{2}}{T_{5}(P)} \\
 & = & \frac{Md_{\mu}(t_{e}+t_{c})}{\frac{Mp}{P}\left(t_{e}+\left(\log_{2}d_{\mu}\right)t_{c}\right)+t_{i}+t_{s}(M)} \\
 & = & \frac{P}{\frac{p\left(t_{e}+\left(\log_{2}d_{\mu}\right)t_{c}\right)}{d_{\mu}(t_{e}+t_{c})}+\frac{P\left(t_{i}+t_{s}(M)\right)}{Md_{\mu}(t_{e}+t_{c})}}
\end{eqnarray*}

\noindent with efficiency

\begin{eqnarray*}
E_{5}(P)=\frac{S_{5}(P)}{P} & \approx & \frac{1}{\frac{p\left(t_{e}+\left(\log_{2}d_{\mu}\right)t_{c}\right)}{d_{\mu}(t_{e}+t_{c})}+\frac{P\sigma}{d_{\mu}t_{n}}}
\end{eqnarray*}

\noindent Making the same assumptions about $t_{s}(M)$, $t_{i}$, and $\gamma$,
$S_{5}(P)$ simplifies asymptotically to 

\[
S_{5}(P)\approx\frac{P}{\frac{p\left(t_{e}+\left(\log_{2}d_{\mu}\right)t_{c}\right)}{d_{\mu}(t_{e}+t_{c})}+\frac{P\sigma}{d_{\mu}t_{n}}}\]

For the values of $P$ and $d_{\mu}$ we examine, this should not
be very different from $S_{3}(P)$. However, these equations allow
us to examine when $S_{5}(P)>S_{3}(P)$, which occurs when 

\begin{eqnarray*}
\frac{p\left(t_{e}+\left(\log_{2}d_{\mu}\right)t_{c}\right)}{d_{\mu}(t_{e}+t_{c})} & < & 1\text{, \,\,\,\ or}\\
p\left(t_{e}+\left(\log_{2}d_{\mu}\right)t_{c}\right) & < & d_{\mu}(t_{e}+t_{c})\\
p & < & \frac{d_{\mu}(t_{e}+t_{c})}{t_{e}+\left(\log_{2}d_{\mu}\right)t_{c}}\end{eqnarray*}

\noindent If we further assume $t_{e}$ and $t_{c}$ are roughly equivalent
operations (both being comparisons) and each taking time $t,$ we
can simplify this to 

\begin{eqnarray}
p & < & \frac{2td_{\mu}}{t\left(1+\log_{2}d_{\mu}\right)}\nonumber \\
p & < & \frac{2d_{\mu}}{1+\log_{2}d_{\mu}}\label{eq:pBound}\end{eqnarray}

\noindent For practical values of $d_{\mu}$, the slope of the graph of \ref{eq:pBound}
is around $\nicefrac{1}{3}$. Since the number of decision nodes grows
faster than the average depth (at a rate dependent on the balancing
of the tree), we should not expect a great speedup from Procedure
\ref{alg:SpecEval1} for any but the most shallow trees.

\section{Experiments on Parallel Classification Tree Algorithms}

The preceding analysis assumes each parallel node execution is independent
from the others. In GPUs, particularly CUDA architecture, this is
not the case. We expect to see a performance benefit due to local
caching of neighboring records read from global memory in bursts,
the SIMD coupling of execution nodes evaluated in parallel for each
sample, having multiple SIMD groups resident and quickly switched
to on the chosen hardware, and other such concerns. These are not
general concerns but are specific to a particular hardware architecture.
In this setting, it makes sense to pursue more specific analysis by
experimentation. The following sections detail experiments done on
the CUDA platform with runtime as the metric of performance.

\subsection{Problem Selection}

We selected the Image Segmentation dataset from UC Irvine's Machine
Learning Repository \cite{UCIImgSeg} as an evaluation problem representative
of tasks in medical imaging or computer vision. This data set consists
of 2310 records for training and an additional 2099 for testing. Each
record consists of 19 real-valued attributes of a $3\times3$ pixel
neighborhood and corresponds to one of 7 discrete classes.

\begin{figure*}
   \centering%
      \includegraphics[bb=0bp 0bp 470bp 453bp,clip,scale=0.9]{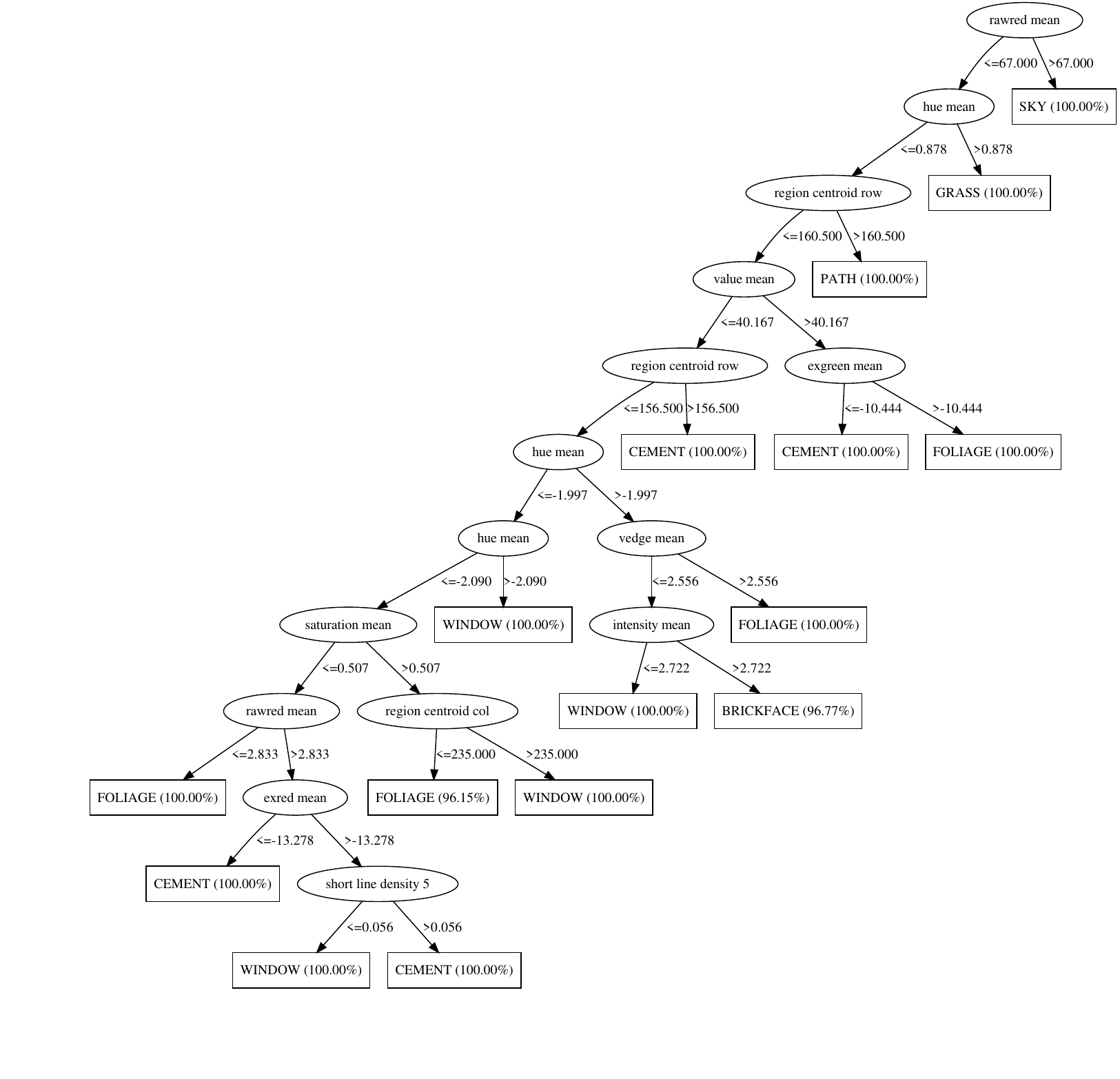}
      \par
   \caption{\label{fig:Experimental-Classification-Tree}Experimental Classification
Tree}
\end{figure*}

To generate a classifier based on this dataset, we used the Orange
component-based machine learning library available from \cite{Orange}.
This library provides Python bindings to a mature C++ machine learning
library. We wrote a Python script to read the training set, train
a classification tree, and generate C++ source code which encodes
that tree according to Procedure \ref{alg:BreadthFirstEnc}. The resulting
tree is shown schematically in Figure \ref{fig:Experimental-Classification-Tree}.
This tree has $N=31$ nodes, 16 leaves, and a depth of 11.

Further, the script also combined the training set and the test set
of records into a single table, then repeatedly randomized and output
the records as C++ source code for easy inclusion in our test program.
This process was repeated until 16,384 C++ records were generated.
This set can be duplicated four times at runtime to create a dataset
having 65,536 records, representing an image of $256\times256$ pixels.

\subsection{Experiment Setup}

\subsubsection{Machine Configuration}

Experiments were performed on a Dell Optiplex 780 with an Intel Core2
Duo E8600 CPU running at 3.33 GHz, 4 GB RAM, and the Windows 7 64-bit
operating system. An NVIDIA Quadro 2000 GPU card was added with 1
GB of 128-bit RAM with a bandwidth of 41.6 GB/s and 192 CUDA cores
in 4 SMs of 48 cores each with a 1.25 GHz processor clock. Software
on the system included the NVIDIA driver version 263.06 and the CUDA
3.2.1 runtime Dll version 8.17.12.6303. All compilation was done with
Microsoft Visual Studio 2008 and the CUDA 3.2 Development Toolkit,
with project files generated by CMake version 2.8.3.

\subsubsection{Tests Conducted}

We created a program which, after building a dataset of 65,536 records,
ran three tree evaluation functions 500 times each on the full dataset.
For each function call, the Windows high performance counter was started
before and stopped after the call and the delta time was accumulated.
This is called the \emph{outer time} for the algorithm. For those
functions using a CUDA kernel, a similar \emph{inner time} was collected
around just the kernel invocation and excluded any time for memory
copy to or from the GPU. During the kernel runtime, the host CPU was
made to wait until the kernel completed. The three functions evaluated
were as follows:

\begin{description}
\item [{\texttt{EvalTree()}:}] This function implements Procedure \ref{alg:SerialEval},
a serial algorithm running on the host. Note that this function records
no inner time and that the outer time does not include any memory
copies since none are required for the host to evaluate the dataset.
\item [{\texttt{EvalTreeBySample()}:}] This is the data parallel algorithm
given in Procedure \ref{alg:DataEval}. This function is written in
CUDA C, and performs a host-to-device copy of the dataset and the
tree definition before invoking the kernel. The grid is formed of
512 blocks having 128 threads each, all single-dimensioned. Only one
record is evaluated per thread (i.e. $m=1$.) For this function (and
all other CUDA functions), the tree is copied to device constant memory
for caching purposes. When the kernel completes, the host copies the
resulting class assignments back to host memory and frees all device
resources.
\item [{\texttt{EvalTreeByNode()}:}] This function fully implements the
improved speculative algorithm corresponding to Procedure \ref{alg:SpecEval1}
with the following considerations: constant memory is used for the
processor-node map and static leaf path buffers in addition to the
tree definition; multiple reductions (specifically 2, determined empirically)
are performed per iteration of the $path$ reduction loop; and the
explicit \textbf{barrier}() operations are omitted since each thread
executes synchronously within a warp. The shared memory $path$ variable
is initialized from the static leaf buffer only once at kernel invocation.
This is safe since leaves never change and internal nodes are re-initialized
by the node evaluation step done for each record. The grid is set
to 128 blocks of $16\times16$ threads. Thus each block processes
16 record groups in parallel, each record group using $p=16$ threads
(a half-warp) to evaluate a record. Note that there are only 15 internal
nodes in the tree, so one thread is idle per record group (assigned
to a phantom node). With $128\times16$ record groups, each must process
$m=32$ records per group to cover 65,536 records exactly. Having
thread geometry exactly match data size allows us to remove checks
for over-sized grids--a non-portable practice but one with a noticeable
performance effect. Data copies to and from the device were the same
as in \texttt{EvalTreeBySample()}.
\end{description}

After each CUDA function call, the returned buffer of class assignments
was compared to the results obtained using the serial algorithm, and
any discrepancies were reported. None were found.

The entire program also ran with the CUDA profiler enabled. This facility
captures device timestamps and other metrics resulting from the program
execution.

\subsection{Results}

The program output giving the outer and inner times along with related
statistics is summarized in Table \ref{tab:OuterInnerTimes}. Most
notable is that the serial evaluation on the host is twice as fast
as the fastest parallel GPU version. This is surprising but perhaps
a bit misleading, since no great pains were taken to optimize the
memory copy tasks, all done in series. Pinning and aligning the host
memory buffers and overlapping copies with computation are viable
techniques to boost performance for this problem. However, it does
point out that the methods used in \cite{Sharp08} by Sharp to measure
a speedup of two orders of magnitude may be mismatched with our methods.
Sharp also does not give the serial algorithm used to compare with
the parallel algorithm, suggesting that perhaps a branchless serial
algorithm performs better than that used in \cite{Sharp08}.

\noindent %
\begin{table*}[t]
\centering%

\caption{\label{tab:OuterInnerTimes}Outer and Inner Times According to High-Performance
Counter}

\resizebox{0.75\textwidth}{!}{%

\begin{tabular}{>{\centering}m{1.8cm}>{\centering}m{1.1cm}>{\centering}m{1.1cm}>{\centering}m{1.1cm}>{\centering}m{1.1cm}>{\centering}m{1.1cm}>{\centering}m{1.1cm}>{\centering}m{1.1cm}>{\centering}m{1.1cm}}
\toprule 
{\footnotesize Algorithm } & {\footnotesize Average Outer Time }{\footnotesize \par}

{\footnotesize ($\mu s$)} & {\footnotesize Min Outer Time ($\mu s$) } & {\footnotesize Max Outer Time}{\footnotesize \par}

{\footnotesize ($\mu s$)} & {\footnotesize Std Dev } & {\footnotesize Average Inner Time ($\mu s$)} & {\footnotesize Min Inner Time}{\footnotesize \par}

{\footnotesize ($\mu s$)} & {\footnotesize Max Inner Time ($\mu s$)} & {\footnotesize Std Dev }\tabularnewline
\midrule 
{\footnotesize EvalTree (Host) } & \textbf{\footnotesize 1914.16} & {\footnotesize 1900.48} & {\footnotesize 2343.65} & {\footnotesize 43.481} & {\footnotesize N/A} & {\footnotesize N/A} & {\footnotesize N/A} & {\footnotesize N/A}\tabularnewline
\midrule 
{\footnotesize EvalTreeBy}{\footnotesize \par}

{\footnotesize Sample } & \textbf{\footnotesize 3907.57} & {\footnotesize 3794.19} & {\footnotesize 4741.2} & {\footnotesize 77.2049} & \textbf{\footnotesize 538.235} & {\footnotesize 525.705} & {\footnotesize 769.309} & {\footnotesize 15.3554}\tabularnewline
\midrule 
{\footnotesize EvalTreeBy}{\footnotesize \par}

{\footnotesize Node } & \textbf{\footnotesize 3785.29} & {\footnotesize 3685.17} & {\footnotesize 4677.76} & {\footnotesize 87.0612} & \textbf{\footnotesize 404.466} & {\footnotesize 394.817} & {\footnotesize 432.698} & {\footnotesize 10.9616}\tabularnewline
\bottomrule
\end{tabular}

}

\end{table*}

In our main result, comparing the inner times for kernel execution
we see a roughly 25\% performance increase in \texttt{EvalTreeByNode}
over \texttt{EvalTreeBySample}. Further experiments on \texttt{EvalTreeByNode}
showed that inclusion of a conditional for checking an over-sized
warp increased runtime to roughly the same as \texttt{EvalTreeBySample}.
With $m=1$, timings were again roughly equal, showing that the expense
of the initial load of static paths and the processor-node map are
amortized over multiple record iterations. Values for $m>32$ (with
related block resizing) showed no significant benefit. This and other
experiments suggests that CUDA thread scheduling is as efficient as
iterating in a for loop.

\begin{figure*}[t]
   \centering
      \resizebox{0.75\textwidth}{!}{%
         \includegraphics{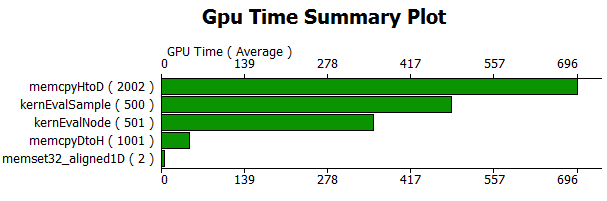}
      }
   \caption{\label{fig:Average-timings-taken}Average timings taken by CUDA runtime
    over 500 executions ($\mu s$)}
\end{figure*}

Examination of the CUDA profiler output shows similar results for
kernel timings (Figure \ref{fig:Average-timings-taken}), though uniformly
lower than those measurable outside of the CUDA driver. The GPU times
confirm a \textasciitilde{}25\% improvement in kernel times of $353.47\mu s$
vs $485.17\mu s$. The time in the graph for {}``memcpyHtoD'' shows
the copy time of the data set and tree definitions (two invocations
per execution) for both CUDA functions over 500 iterations each. Adding
this and the {}``memcpyDtoH'' time to each of the kernel times gives
the outer time for each function, less time taken by the host to allocate/free
buffers and manage the function calls.

The profiler data also shows \texttt{EvalTreeByNode} taking an average
of 4373 divergent branches across all threads due to the half-warp
scheduling, whereas \texttt{EvalTreeBySample }shows 3530 across all
threads, as each thread in a warp will iterate through the tree a
different number of times. \texttt{EvalTreeByNode} had a global cache
read hit rate of 70\%, while \texttt{EvalTreeBySample} had a hit rate
of only 31\%.

With fewer threads per block, \texttt{EvalTreeBySample} encounters
the limit on active blocks, leaving the achieved occupancy rate at
66\%. \texttt{EvalTreeByNode} avoids this issue and achieves 100\%
occupancy. This increases the number of global memory requests for
record data that can be active, and thus enhances the effect of latency
hiding by the warp scheduler. This can be seen in the global memory
write throughput of 0.643 GB/s versus 4.68 GB/s. Read throughputs
are roughly equal at 14 GB/s (due to caching), giving overall global
memory throughputs of 15.43 GB/s for \texttt{EvalTreeBySample} and
19.41 GB/s for \texttt{EvalTreeByNode}.

\section{Conclusion}

We have shown a speculative decomposition algorithm for parallel classification
tree evaluation that surpasses the performance of a data decomposition
parallel algorithm on the CUDA platform. When ignoring the common,
serial algorithm setup processing, the speculative approach is 25\%
faster than the data parallel approach in our particular problem instance.
This demonstrates how different parallel decomposition techniques
can maximize the advantages of a given platform. In a SIMD environment,
we see that speculative decomposition into many time-uniform tasks
can have a helpful effect even at the cost of less efficient hardware
utilization. We also see a good example of implementation results
deviating from asymptotic theoretical analysis. This is most true
when fundamental assumptions, such as independent execution units,
do not hold in the implementation as is the case here. Ultimately,
the best performance requires a careful balance of machine and algorithm
for a specific problem.

Additionally, we've seen that measurement techniques which do not
include the entire program overhead of distributing data or that compare
different algorithms can lead to confusing results. Though we have
implemented a very similar program to \cite{Sharp08}, our serial
host implementation is roughly twice as fast when all overhead in
included, compared to 100 times faster as Sharp reports. Surely, some
difference in host speed, GPU power, and lower overhead cost when
processing forests rather than single trees is responsible for part
of this discrepancy. The remaining difference suggests that the branchless
evaluation algorithm ought to be used as the best known serial algorithm
for speedup comparisons.

\section{Further Work}

The breadth of this result should be tested against other tree geometries
(e.g. more or less balanced, deeper or more shallow) and record distributions
(ordered vs. random) to observe the effect different data organizations
can have on run times. Also, application of these algorithms to more
traditional SIMD, i.e. vector, processors would be interesting. Comparing
CUDA compute 1.x devices with 2.x devices might also provide additional
insights.

To extend the current work, application to very large trees might
be achieved by evaluating only a small {}``window'' on the tree,
starting at a root node and evaluating only the next few levels. Once
reduced, the resulting node would then become the root of the next
window and the process repeated. This approach may be useful in overcoming
SIMD concurrency limits (such as on a vectored processor) or the exponential
growth of memory demand for deeper and deeper levels of the tree.

\bibliographystyle{plain}
\bibliography{SpecTreeEvalPrePrint}

\begin{thebibliography}{10}

\bibitem{Baumstarck}
Paul Baumstarck.
\newblock {GPU} parallel processing for fast robotic perception.
\newblock {Thesis, Engineer's degree}, Stanford University, December 2009.

\bibitem{Ben-Haim2010}
Yael Ben-Haim and Elad Tom-Tov.
\newblock A streaming parallel decision tree algorithm.
\newblock {\em J. Mach. Learn. Res.}, 11:849--872, March 2010.

\bibitem{CudaZone}
NVIDIA Corporation.
\newblock {CUDA Zone}.
\newblock \url{http://www.nvidia.com/object/cuda_home_new.html}, Feb 2011.

\bibitem{DevZone}
NVIDIA Corporation.
\newblock {NVIDIA Developer Zone}.
\newblock \url{http://developer.nvidia.com/object/gpucomputing.html}, Feb 2011.

\bibitem{Stair}
Stephen Gould, Olga Russakovsky, Ian Goodfellow, Paul Baumstarck, Andrew~Y. Ng,
  and Daphne Koller.
\newblock {The STAIR Vision Library (v2.4)}.
\newblock \url{http://ai.stanford.edu/~sgould/svl}, May 2010.

\bibitem{Jin2002}
Ruoming Jin and Gagan Agrawal.
\newblock Shared memory parallelization of decision tree construction using a
  general data mining middleware.
\newblock In {\em Proceedings of the 8th International Euro-Par Conference on
  Parallel Processing}, Euro-Par '02, pages 346--354, London, UK, 2002.
  Springer-Verlag.

\bibitem{Scalparc}
Mahesh~V. Joshi, George Karypis, and Vipin Kumar.
\newblock Scalparc: A new scalable and efficient parallel classification
  algorithm for mining large datasets.
\newblock In {\em Proc. of the International Parallel Processing Symposium},
  pages 573--579, 1998.

\bibitem{Kirk10}
David~B. Kirk and Wen-mei~W. Hwu.
\newblock {\em {Programming Massively Parallel Processors: A Hands-on
  Approach}}.
\newblock Morgan Kaufmann, 1st edition, Feb 2010.

\bibitem{Orange}
Faculty of~Computer Laboratory~of Artificial~Intelligence and Information
  Science.
\newblock Orange for python 2.6.
\newblock \url{http://orange.biolab.si/}.

\bibitem{Sliq}
Manish Mehta, Rakesh Agrawal, and Jorma Rissanen.
\newblock Sliq: A fast scalable classifier for data mining.
\newblock In {\em Proc. of the Fifth International Conference on Extending
  Database Technology (EDBT)}, pages 18--32, Avignon, France, March 1996.

\bibitem{Nair02}
A.~Nair, B.~Kuban, E.~Tuzcu, P.~Schoenhagen, S.~Nissen, and D.~Vince.
\newblock Coronary plaque classification with intravascular ultrasound
  radiofrequency data analysis.
\newblock {\em Circulation}, 106:2200--2206, October 2002.

\bibitem{Oliver05}
Arnau Oliver and Jordi Freixenet.
\newblock Automatic classification of breast density.
\newblock In {\em IEEE International Conference on Image Processing}, pages
  1258--1261, 2005.

\bibitem{Sanders2010CUDA}
Jason Sanders and Edward Kandrot.
\newblock {\em {CUDA by Example: An Introduction to General-Purpose GPU
  Programming}}.
\newblock Addison-Wesley Professional, 1st edition, July 2010.

\bibitem{Sprint}
John Shafer, Rakeeh Agrawal, and Manish Mehta.
\newblock Sprint: A scalable parallel classifier for data mining.
\newblock In {\em Proceedings of the 22nd International Conference on Very
  Large Databases (VLDB)}, pages 544--555. Morgan Kaufmann, September 1996.

\bibitem{Sharp08}
Toby Sharp.
\newblock Implementing decision trees and forests on a gpu.
\newblock In {\em European Conference on Computer Vision (ECCV) 2008}, volume
  5305 of {\em Lecture Notes in Computer Science}, pages 595--608. Springer,
  2008.

\bibitem{GpuMl}
D.~Steinkraus, I.~Buck, and P.Y. Simard.
\newblock Using {GPUs} for machine learning algorithms.
\newblock In {\em Document Analysis and Recognition, 2005. Proceedings. Eighth
  International Conference on}, pages 1115 -- 1120 Vol. 2, 29 Aug.-1 Sept.
  2005.

\bibitem{UCIImgSeg}
{UCI Machine Learning Repository}.
\newblock {Image Segmentation data set}.
\newblock \url{http://archive.ics.uci.edu/ml/datasets/Image+Segmentation},
  November 1990.

\bibitem{Zaki98}
Mohammed~J. Zaki, Ching-Tien Ho, and Rakesh Agrawal.
\newblock Parallel classification for data mining on shared-memory
  multiprocessors.
\newblock {\em Data Engineering, International Conference on}, 0:198, 1999.

\end{thebibliography}

\end{multicols}

\end{document}